\documentclass[11pt]{article}

\usepackage{amsfonts}
\usepackage{amsbsy} 
\usepackage{epsfig}
\usepackage{latexsym}
\input amssym.def
\input amssym.tex

\def\bequ{\begin{equation}}
\def\eequ{\end{equation}}
\def\barr{\begin{array}}
\def\earr{\end{array}}

\def\ben{\begin{equation}}
\def\een{\end{equation}}
\def\bena{\begin{eqnarray}}
\def\eena{\end{eqnarray}}

\newcommand{\sect}[1]{\setcounter{equation}{0}\section{#1}}

\advance \topmargin by -\headheight
\advance \topmargin by -\headsep
\evensidemargin \oddsidemargin
\advance\hoffset by -3mm  
\advance\voffset by  8mm  

\def\const{\rm constant}

\def\bR{\Bbb R}

\def\bE{\Bbb E}

\def\b1{e^0}

\begin{document}
\hfuzz=100pt
\title{{\Large \bf{Born-Infeld Theory and Stringy Causality}}} 
\author{G W Gibbons\footnote{Email address: gwg1@damtp.cam.ac.uk} \ \& C A R Herdeiro \footnote{Email address: car26@damtp.cam.ac.uk}
\\ (D.A.M.T.P.)
\\Department of Applied Mathematics and Theoretical Physics,
\\ Centre for Mathematical Sciences,
\\ University of Cambridge, 
\\ Wilberforce Road,
\\ Cambridge CB3 0WA,
 \\ U.K.}

\date{August 2000}

\maketitle
\centerline{DAMTP-2000-71}

\begin{abstract} 
Fluctuations around a non-trivial solution of Born-Infeld theory have a limiting speed given not by the Einstein metric but the Boillat metric. The Boillat metric is S-duality invariant and conformal to the open string metric. It also governs the propagation of scalars and spinors in Born-Infeld theory. We discuss the potential clash between causality determined by the closed string and open string light cones and find that the latter never lie outside the former. Both cones touch along the principal null directions of the background Born-Infeld field. We consider black hole solutions in situations in which the distinction between bulk and brane is not sharp such as space-filling branes and find that the location of the event horizon and the thermodynamic properties do not depend on whether one uses the closed or open string metric. Analogous statements hold in the more general context of non-linear electrodynamics or effective quantum-corrected metrics. We show how Born-Infeld action to second order might be obtained from higher-curvature gravity in Kaluza-Klein theory. Finally we point out some intriguing analogies with Einstein-Schr\"odinger theory.
\end{abstract}

04.50.+h; 04.70.-s; 11.25.-w; 11.10.Lm

\sect{Introduction}

A striking feature of much recent work on open string
states in String/M-Theory is the considerable insights
afforded by the Born-Infeld \cite{BI} approximation.
Reciprocally, String/M-Theory has provided a rationale
for some of the hitherto mysterious and only partially understood
properties of this remarkable theory.

The existence of a limiting electric field strength, which 
was originally the {\it raison d'etre} of Born-Infeld
theory, now finds a dynamical justification
in the increasingly copious 
production of electrically  charged open string states
as one approaches the critical value \cite{porratti}.   
More subtly, the electric-magnetic duality symmetry
of Born-Infeld theory, a non-linear generalization of Hodge
duality first recognized by Schr\"odinger \cite{S1}, 
may be viewed as a special case of S-duality. 
In fact electric-magnetic duality
is a special case of Born-Reciprocity \cite{B}, a transformation 
which acts as a rotation in phase space
$
(p,q) \rightarrow 
(p \cos \theta + q \sin \theta, q \cos \theta- p\sin \theta)
$. In non-linear electrodynamics, the phase space variables might be considered to be 
$\bf B$ and ${\bf D}= \partial L / \partial {\bf E}$, which are canonically
conjugate variables in the sense of the Poisson bracket\footnote{We use the convention $\{x^i,p_j\}_{P.B.}=\delta^i_j$.}
\ben
\Bigl \{B_i({\bf x}), D_j({\bf y}) \Bigr \}_{P.B} =-\epsilon_{ijk}\partial_k \delta({\bf x}-{\bf y}).
\label{canbd}
\een
Born reciprocity applied to string theory
gives rise to T-duality \cite{V}. According to Hull and Townsend \cite{hull} 
T and S duality are included in the more general U-duality symmetry.
This leads naturally to the question of whether
Born-Infeld theory is the only
non-linear electrodynamic theory admitting electric-magnetic duality.
It is not \cite{GR}.

Another striking feature of Born-Infeld theory, 
is that it admits BIon solutions. These are exact solutions of the full
non-linear theory with distributional sources with finite total
energy. They can now be understood in terms of strings ending
on D-branes \cite{G, CM}. 

Schr\"odinger
recognized yet another remarkable property of Born-Infeld theory:
 viewed as a non-linear  optical theory,  
Born-Infeld theory exhibits uncommon properties with respect 
to the scattering of light by light. He constructed exact wave like
solutions of the full non-linear equations
representing light pulses with
solitonic properties. They  pass through one another without
scattering \cite{S2,S3}. This can be also understood from a string theory
point of view \cite{G}.

Some time after  Shr\"odinger's work
it was realized that the propagation
of Born-Infeld fluctuations around a background solution
has exceptional causal
properties \cite{B13,P} and that the theory 
 also admits exact solutions
exhibiting these exceptional  properties \cite{BB}. From the String Theory perspective, this relates to the recent interest in open string theory 
in a constant background Kalb-Ramond potential $B_{\mu \nu}$
and thus with gauge theory in a flat non-commutative
spacetime \cite{seiwit}. The  Kalb-Ramond potential
$B_{\mu \nu}$ appears in the Born-Infeld action
 in the combination $F_{\mu \nu} + B_{\mu \nu}$
and so from the Born-Infeld point of view, a constant $B_{\mu \nu}$ field
may be regarded as a background solution of Born-Infeld theory.
Some of the open string states propagating around the constant background
$B_{\mu \nu}$  field may be identified (at least in the abelian case)
 with fluctuations of the Born-Infeld theory.
Thus we need to understand the causal structure of their
propagation. It turns out that this is governed by a metric,
 $G_{\mu \nu}=g_{\mu \nu} -B_{\mu \alpha} g^{\alpha \beta } B_{\beta \nu}$\footnote{$B_{\mu \nu}$ may be taken to stand for the background field
or for the constant Kalb-Ramond 2-form. In string language we are using units
in which $2 \pi \alpha ^\prime=1$ }, which {\sl differs}
 from the usual spacetime metric
 $g_{\mu \nu}$ \cite{seiwit}. In fact we have two light-cones: the usual
light-cone given by $g_{\mu \nu}$ which  governs the
propagation of closed string states such as the graviton
and that given by 
$G_{\mu \nu}$ which governs the propagation of open string states such as the Born-Infeld photon. In general, the former lies outside the latter
except in two privileged directions corresponding
to the two principal null directions or eigenvectors
of the background two-form field. To put things provocatively,
gravitons almost always travel faster than light. 

One immediate
consequence is that if the closed string metric admits no closed
timelike curves then neither will the open string metric.
Another obvious consequence is that if the closed string metric contains
an event horizon   then the open string metric will
also contain an event horizon which lies inside or 
on the closed string event horizon.

The comments in the last two paragraphs encode the global theme of this paper. Within many physical theories, two non-conformal metric structures arise, 
\bequ
g_{\mu \nu} \ \ \ \ \  and  \ \ \ \ \ g_{\mu \nu} + S_{\mu \nu},
\label{bimet}
\eequ
where $S_{\mu \nu}$ is some symmetric two-tensor. Geometrically they represent two sets of light-cones. Questions regarding causality or the existence of event horizons might then become particularly subtle. It is our purpose to discuss such issues in several contexts. For the aforementioned open versus closed string theory causal structures,  there is some advantage in  placing these
properties  in the general context of non-linear
electrodynamic theories, just as in the case of electric-magnetic
duality. Particularly so because
a quite separate strand of recent research
has been concerned with analogues of black holes, closed timelike
curves and circular null-geodesics in  non-linear electrodynamics \cite{novello}. There is also   considerable interest
in black holes in theories of non-linear electrodynamics coupled
to Einstein gravity (some recent references are \cite{beato}).

In section 2 we look at general non-linear electrodynamics in a four-dimensional flat background. Hence $g_{\mu \nu}$ is the Minkowski metric. The study of the propagation of fluctuations of the electromagnetic field in a given electromagnetic background introduces naturally a second metric structure: the Boillat metric $A_{\mu \nu}^{Boillat}$. We emphasize the special properties of Born-Infeld theory in at least four senses: the existence of both electric-magnetic and Legendre duality and the absence of both bi-refringence and shocks.

In section 3, we specialize to the case when the electrodynamics theory is Born-Infeld. One can deal with a general curved background. So the natural background geometry is described by $g_{\mu \nu}$, the closed string metric. But open strings propagating in a non-trivial electromagnetic field or Kalb-Ramond potential see a different metric: the open string metric $G_{\mu \nu}$, conformal to $A_{\mu \nu}^{Boillat}$. We shall then see that if the closed string metric is static and the
Born-Infeld field is pure electric or pure magnetic
then the open string metric cannot have a non-singular
event horizon distinct from the one given by the closed string metric, because on it the electric field $\bf E$ must either equal its limiting value
or the magnetic field $\bf B$ must diverge.  Note that the metric $G_{\mu \nu}$ is not invariant, even up to a conformal
factor, under Hodge duality $\delta B_{\mu \nu}=\star B_{\mu \nu}$
but, as we shall see, it {\sl is}  invariant up to a conformal factor under electric-magnetic duality rotations.

In section 4 we show that scalar and spinor fluctuations around a Born-Infeld background are governed by the open string metric.

In section 5 the bi-metric theme takes another perspective. Recently \cite{her} examples have been given of how dimensional reduction can alter the causal structure of stringy black holes. Considering a trivial dilaton field, the relation between the lower dimensional metric $g_{\mu \nu}$ and the higher dimensional one $\hat{g}_{\mu \nu}$ is of the form (\ref{bimet}), with $\hat{g}_{\mu \nu}=g_{\mu \nu}+A_{\mu}A_{\nu}$. With this motivation, we look to higher-order Kaluza-Klein theory.  We notice that it is possible to obtain Born-Infeld theory to second order and still avoid ghosts, as long as the higher dimensional graviton is only excited along the compact dimensions. We similarly show that the effective theory for QED, the Euler-Heisenberg theory, may be obtained in this fashion, and discuss some properties of the theory obtained by starting with an Einstein-Hilbert plus Gauss-Bonnet action in higher dimensions \cite{lemos}. This gives an application of the general concepts discussed in section 2.

In section 6 we start by reviewing the results of section 2 considering the gravitational effects of the electromagnetic background field. One is then led to consider besides the usual Einstein metric $g_{\mu \nu}$ an effective co-metric of the form $g^{\mu \nu}+AR^{\mu \nu}$. Another possible origin for such effective metric is quantum renormalization of the propagator of test fields in a fixed background. We then discuss the universality of black holes event horizons and thermodynamic properties, by applying a result derived in the context of quantum renormalized metrics to the case of the Boillat metric for non-linear electrodynamics coupled to gravity.

In section 7, we review an old attempt of Einstein and Schr\"odinger to construct a unified theory of gravity and electromagnetism (see \cite{deser} for original references). One then introduces a metric which has an antisymmetric part. The symmetric part $g_{\mu \nu}$ and the inverse of the symmetric part of the inverse of the full metric, $A_{\mu \nu}^{\rm Eins-Schro}$ have remarkable similarities with the closed and open string metric, as first noticed by Boillat \cite{B11}. We discuss some exact solutions found by Papapetrou \cite{papape}.

We close with a discussion.

\sect{Causality in Non-Linear Electrodynamics}

\subsection{Characteristics and effective geometry}

We consider a general Lagrangian $L=L(x,y)$  
depending on the Lorentz invariants $x={ 1\over4}  F_{\mu \nu} F^{\mu \nu}$
and $y={ 1\over4}  F_{\mu \nu} \star F^{\mu \nu}$ .\footnote{We use a mainly minus metric signature and, contrary to Boillat and some other references
who use the opposite sign we choose $L$ to have the standard sign such that for Maxwell theory $L=-x$. Subscripts indicate partial differentiation.} These are the only independent Lorentz invariants in four spacetime dimensions. 
The energy momentum tensor is given by      
\ben
T_{\mu \nu}= -L_x T^{\rm Maxwell}_{\mu \nu}  + { 1\over 4} T g_{\mu \nu},
\een
where the trace and the Maxwell energy-momentum tensor are given by
\ben
T\equiv T^\mu_\mu =-4(L-xL_x-yL_y), \ \ \ \ \ \ \ \ \ \ \ \
T^{\rm Maxwell}_{\mu \nu}= -F_{\mu \alpha} F_{\nu \beta} g^{\alpha \beta}+x g_{\mu \nu}.
\een

Since both $ T^{\rm Maxwell}_{\mu \nu}$ and $g_{\mu \nu}$ (with mainly minus signature) 
satisfy the dominant energy condition, and the set   
of energy momentum tensors
satisfying the dominant energy condition is  a convex cone, a sufficient requirement 
for $T_{\mu\nu}$ to satisfy the dominant energy condition is that 
$L_x <0$ and  $T\ge0$.  An argument of Hawking and Ellis \cite{hawell} then shows that 
propagation in the full non-linear theory is causal in the sense that if at time zero all fields vanish
outside some compact set,
then they will vanish outside the future of that set.
In general one expects the fields to advance into empty space with no background field 
 at the speed of light and this expectation is supported by the observation
(originally due to Schr\"{o}dinger \cite{S2}) 
that any solution of Maxwell's linear electrodynamics
 with vanishing invariants,   
$x=y=0$,  will also be an exact solution of non-linear electrodynamics.
Among such so-called self-conjugate solutions are the usual
plane wave solutions which have unit speed.  

If a background field is present however these arguments require 
re-examination. One approach might be to look at the energy momentum tensor
of the fluctuations. We shall not do this here but begin by considering 
the {\it characteristics},
 which by definition are hypersurfaces
along  which {\sl weak} discontinuities
propagate. Assuming $F_{\mu \nu}$ to be discontinuous across the  $S(x^\mu)= {\rm constant }$ the characteristics are given by \cite{B6,B13,P,GDP}
\ben
(T_{\rm Maxwell}^ {\mu \nu}+\mu g^{\mu \nu})\partial _\mu S \partial _\nu S =0.
\label{charac}
\een
This has the form of a relativistic Hamilton-Jacobi equation for massless particles with effective co-metric $T_{\rm Maxwell}^ {\mu \nu}+\mu g^{\mu \nu}$, and where $S$ would be the action function. This effective metric also governs the propagation of 
weak, but not necessarily discontinuous fluctuations
around a background. Later we will turn
to the propagation of shocks and the behavior
of fully non-linear fluctuations. The function $\mu=\mu(x,y)$ satisfies
\ben
\varpi \mu^2 +\mu +\omega -\varpi (x^2+y^2)=0,
\label{quadra}
\een
where,
\ben
\varpi ={ L_{xx}L_{yy}-L_{xy}^2 \over L_{x}(L_{xx}+L_{yy})},
\een
and 
\ben
\omega={L_x+x(L_{xx}-L_{yy})+2yL_{xy} \over L_{xx}+L_{yy} } . 
\een

In the general case
the characteristics exhibits bi-refrigence: $\mu(x,y)$, which for convenience we parameterize as $\mu(x,y)= x + \zeta_\pm (x,y),
$ can take {\sl two values}, depending upon the polarization state and the background field. Thus there are, in general, two metrics.
The interpretation of the quantities $\zeta_\pm$ is that they
correspond to critical electric field strengths above which
the theory breaks down. For exceptional
theories the two values of $\zeta_\pm$ coincide and there
is a single light-cone  and no bi-refringence. Exceptional theories fall into two classes. The first has $\varpi=0$. This happens for instance
if the Lagrangian $L$ is independent of $y$, which includes Maxwell's theory as a special case. But not all theories with $\varpi=0$ are exceptional in this sense. In fact, although (\ref{quadra}) still encodes relevant information for this case, it does not contain \textit{all} the information anymore. One example will be given in section 5.

For $\varpi \neq 0$, the only exceptional theory is Born-Infeld \cite{B13}. The latter is also very special in that $\zeta_\pm$ is a constant
independent of $x$ and $y$. It is the only theory for which this is true.
We shall use  units in which this constant
is taken to be one. 

The condition that the theory admit electric-magnetic duality rotations
is rather weaker. It suffices that ${\bf B}\cdot {\bf E}={\bf D}\cdot {\bf H}$ \cite{GR}, which implies that the Lagrangian satisfy the  first order Hamilton-Jacobi type equation
\ben
y(L_x^2-L_y^2)-2xL_{x}L_{y}=y.
\een
 
 The characteristics or wave surfaces
may be thought of as null
hypersurfaces of a metric whose null geodesics 
correspond to  the {\it rays}.  Note that the characteristics and the rays
depend only a conformal equivalence class of metrics, defined by (\ref{charac}). A particular
choice of conformal representative used by Boillat, which
we shall refer to as the Boillat metric and co-metric,
is given by 
\ben
A^{\rm Boillat}_{\mu \nu}= { 1\over \sqrt{\mu^2-x^2-y^2}} 
\bigl (\mu g_{\mu \nu}
-T_{\mu \nu} ^ {\rm Maxwell} \bigr ) 
\label{boime}
\een
\ben
C_{\rm Boillat} ^{\mu \nu}= { 1\over \sqrt{\mu^2-x^2-y^2}}
 \bigl (
\mu g^{\mu \nu}+T^{\mu \nu} _{\rm Maxwell}  \bigr)  , 
\label{boicome}
\een
so that\footnote{Throughout this paper all indices will 
be raised or lowered using the usual Einstein
or closed string metric $g_{\mu \nu}$ {\sl with the exception}
of the open string metric $G_{\mu \nu}$ whose inverse 
is denoted by 
$G^{\mu \nu}$ in accordance with string theory conventions.}  
$A^{\rm Boillat} _{\mu \alpha} C_{\rm Boillat} ^{\alpha \nu}=\delta ^\nu _\mu$. As we shall see in detail
later, in the case of Born-Infeld theory, the open-string
metric $G_{\mu \nu}$ and the Boillat metric  $A^{\rm Boillat}_{\mu \nu}$
are conformal. 

Because
\ben
A^{\rm Boillat}_{\mu \nu}= { \mu -x \over \sqrt{\mu^2-x^2-y^2}} 
\bigl (g_{\mu \nu} -{ 1 \over \mu -x } F_{\mu \alpha} g^{\alpha \beta} F_{\beta \nu}
\bigr ), 
\een
the Boillat metric has a remarkable expression as a sort of
square root:
\ben
A ^{\rm Boillat}= { \mu-x \over \sqrt{\mu^2-x^2-y^2}} 
(g+ { 1\over \sqrt{\mu-x}} F) g^{-1} (g-  { 1\over \sqrt{\mu-x}}F).
\label{boisqu}
\een
It follows easily that
\ben
\sqrt {-\det A^{\rm Boillat} _{\mu \nu}} =\sqrt{-\det g_{\mu \nu}},
\label{elem}
\een
in other words, the Boillat metric and the spacetime metric induce the 
same volume element. The two principal null vectors common to both cones
are annihilated 
by  $g+ F/ \sqrt{\mu-x} $ or by  $ g- F / \sqrt{\mu-x}$ .

We record for later use that if the background Einstein metric $g$ is flat, then
up to the conformal factor $ 1/ \sqrt{\mu^2-x^2-y^2} $
the Boillat metric is
\ben
(\mu -x) (dt^2-d{\bf x}^2 ) -{\bf E}^2 dt^2 + ({\bf E}\cdot d {\bf x} )^2
\pm 2 {\bf E} \times {\bf B} \cdot d {\bf x} dt -{\bf B}^2 d{\bf x}^2 +({\bf
B} \cdot d{\bf x} )^2.   \label{keymetric}
\een

\begin{figure}
\begin{picture}(0,0)(0,0)
\end{picture}   
\centering\epsfig{file=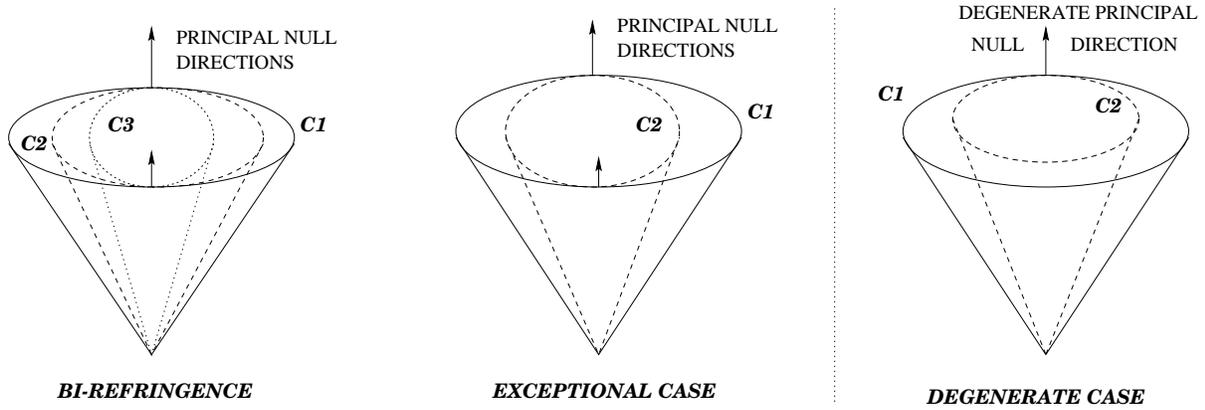,width=16cm}   
\caption{Cones for the Einstein geometry and effective geometry describing the propagation of fluctuations in a non-trivial background $F_{\mu \nu}$ field. If condition (\ref{bradyon}) is obeyed, $C1$ is the Einstein cone, $C2$ (and $C3$ for the non-exceptional case) the effective geometry cones. The cone on the right represents the exceptional degenerate case.}
\label{cones1}
\end{figure}

In the generic case, one may diagonalise the Boillat
metric with respect
to the usual spacetime metric $g_{\mu \nu}$. 
This gives the speeds of propagation of the fluctuations in the associated inertial frame.
In this frame the Poynting vector ${\bf E}\times {\bf H}=
-L_x {\bf E}\times {\bf B}$ vanishes. The  velocities, i.e. the ratio of spacelike to timelike eigenvalues, turn out to be 
\ben
(1,{\mu -\sqrt{x^2+y^2} \over \mu+ \sqrt{x^2 +y^2}},
{\mu -\sqrt{x^2+y^2} \over \mu+ \sqrt{x^2 +y^2}}). 
\label{velo}
\een
Thus in general
there are two directions in which the Boillat-cone touches the usual
Einstein light-cone, corresponding to the first component of (\ref{velo}). These are the principal null directions of 
$F_{\mu \nu}$. Note that $F_{\mu \nu}$ and any duality rotation of it have the same principal null directions. In Figure \ref{cones1}, we represent the light cones for the effective plus Einstein geometry. The left cones illustrate the case with bi-refringence; we then have the Einstein plus two effective geometry cones. For the causal case, the Einstein cone will be $C1$. All the cones touch in two points, along the principal null directions of $F_{\mu \nu}$. The cones in the centre of Figure \ref{cones1} illustrate the exceptional case (like the Born-Infeld or open string theory case) where the effective geometry only possesses one light cone.

It may happen that the two principal null directions coincide. This occurs if and only if $x=0=y$. In this case the metric takes the form,
\bequ
A_{\mu \nu}^{Boillat}=g_{\mu \nu}-l_{\mu}l_{\nu}, 
\eequ
where $l_{\mu}$ is parallel to the principal null direction. The characteristic cone touches the Einstein cone along a single generator. This degenerate case is illustrated for exceptional theories in Figure \ref{cones1} right. For a generic electromagnetic field the principal null directions will coincide on a submanifold of dimension (and also co-dimension) two, $\mathcal{N}$. The complement $\mathcal{M} \setminus \mathcal{N}$ in the spacetime manifold $\mathcal{M}$, may not be simply connected. This gives rise to ambiguities in defining the `complexion' $\frac{1}{2}\arctan{y/x}$ of the electromagnetic field. In many ways, particularly if it is timelike, $\mathcal{N}$ behaves rather like a cosmic string \cite{penrin}.

In string theory, if a dilaton $\Phi$
is present,
one distinguishes between the Einstein metric $g_{\mu \nu}$
and the (closed)-string metric $e^{-2 \Phi} g_{\mu \nu}$. However both
have the same (Einstein) light-cone, i.e., they are conformal. This is because the dilaton
is a state of the closed string. It seems therefore, at least at the level of approximation we are considering, that there are just two causal structures
and two sets of cones: the 
open and the closed. Of course from the strict string theory point of view
one refers  brane and the other to bulk propagation but we have in mind
situations where the distinction is not sharp, such as for example
in the case of space-filling branes, or when considering
gravitons confined to, or at least moving parallel
to, the surface of a brane.

A sufficient condition that the Boillat-cone does not lie outside
the usual Einstein light-cone, i.e. that the speeds never exceed unity is that both $\mu$'s must satisfy 
\ben
\mu >\sqrt{x^2 +y^2}\equiv r.
\label{bradyon}
\een
In terms of the coefficients in (\ref{quadra}) this requirement reads $\omega<-r$, $-1/(2r)<\varpi<0$. Specialized to the Born-Infeld case (\ref{bradyon}) yields positive the quantity under the square root in the Born-Infeld action.

\subsection{Wave surfaces and Ether drift}
Boillat \cite{B13} has calculated the wave-front produced by waves moving
outwards from a point source with respect to an inertial frame
 in which the Poynting vector 
${\bf E}\times {\bf H}=-L_x {\bf E}\times {\bf B}$
 does not vanish. If
\bequ 
w={ 1\over  2} ({\bf E}^2 + {\bf B}^2), \ \ \ a={ \sqrt{\mu+\sqrt{x^2 +y^2} \over \mu+w}}, \ \ \ 
b=\sqrt{ \mu-\sqrt{x^2 +y^2} \over \mu+w },
\eequ
and $c=ab$, so that
$1>a\ge b >c$, he finds that it is given by the family of  ellipsoids (in cartesian coordinates $(x,y,z)$)
\ben
{x^2 \over a^2} + {y^2 \over b^2} + {(z-tv_{\rm drift})^2 \over c^2}= t^2, 
\een
where the drift velocity is given by
\ben
{\bf v}_{\rm drift}= {{\bf E} \times {\bf B} \over \mu +w }.
\label{drift}
\een
Since $v^2_{\rm drift}=(1-a^2)(1-b^2)$, the drift velocity is always
less than one. Therefore, the presence of the background electromagnetic field causes the drift of the origin of disturbances and establishes preferred directions in spacetime; in a sense plays the role of `ether'.

\subsection{Convexity of the Hamiltonian Function}

The energy density or Hamiltonian density
$T_{00}$ should be considered as a function
$H({\bf D},{\bf B})$  of the canonically conjugate variables 
$({\bf D}, {\bf B})$ (in the sense of (\ref{canbd})). Their time evolution is obtained by taking the curl 
of $({\bf H},-{\bf E})$ where the constitutive relation
 $({\bf E}, {\bf H})=
(\partial H / \partial {\bf D} , \partial H / \partial
{\bf B})$ holds.
In other words $({\bf E}, {\bf H})$ and $({\bf D}, {\bf B})$ are related
by a Legendre transformation and in this sense one may regard the variables  $({\bf E}, {\bf H})$ as canonically conjugate to the variables $({\bf D}, {\bf B})$.
Of course this is a  different sense 
of canonically conjugate than that in which $\bf B$ and $\bf D$ are
canonically conjugate. It is a covariant sense in which one thinks
of the space of  Faraday tensors $F_{\mu \nu}$ (possibly subject to the
closure constraint $\partial_{[\mu } F_{\nu \tau]}=0$)
as the covariant
configuration space rather than the non-covariant configuration space
of magnetic induction fields
 $\bf B$ subject to the constraint ${\rm div}{\bf B}=0$.

The Legendre transformation
will be well defined and invertible if and only if the Hamiltonian
density $H({\bf B},{\bf D})$
is a convex function of it's arguments. In other words
the matrix of second derivatives or  Hessian is positive definite.
Note that in general $H$ may be defined
only in a portion of the six-dimensional space of possible ${\bf D}$ and
 $\bf B$ 's and the Legendre transform may only map into
part of  six-dimensional space of possible ${\bf E}$ and
 $\bf H$ 's. Thus for example, in the case of Born-Infeld theory
\ben
H=\sqrt{ (1 + {\bf B}^2)(1 + {\bf D} ^2) -({\bf B} \cdot {\bf D})^2 }-1,
\een
which is, in fact, defined for all $\bf B$ and $\bf D$.
However the inverse Legendre transformation is effected
 by means of the function
\ben
1-\sqrt{ (1-{\bf H} ^2 ) ( 1-{\bf E}^2) - ({\bf E} \cdot {\bf H} )^2 }.
\een
which is defined only over the domain
of   $({\bf E}, {\bf H})$ given by
\ben
{\bf E}^2 + {\bf H}^2 < 1+ ({\bf E}\times {\bf H})^2. 
\een
Born-Infeld theory is one with the same constant upper
bound for both the electric (at zero magnetic field)
and magnetic field  strengths (at zero electric field).

\begin{figure}
\begin{picture}(0,0)(0,0)
\end{picture}   
\centering\epsfig{file=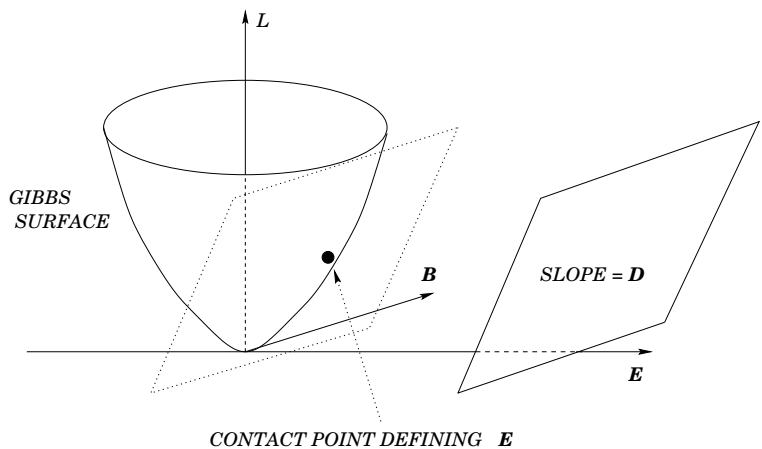,width=10cm}   
\caption{The Gibbs surface is the Lagrangian function. Inverting the constitutive relations, to find ${\bf E}={\bf E}({\bf B}, {\bf D})$, corresponds to finding the ${\bf E}$ coordinate of the contact point  of the Gibbs surface with a plane with slope ${\bf D}$ along a line of constant ${\bf B}$. Convexity is necessary for the inverse constitutive relations to be well defined.}
\label{gibbs}
\end{figure}

Of course it should be borne in mind that
singling out a particular pair of variables 
is rather artificial. The underlying invariant
geometric structure is
the 12-dimensional  symplectic vector space 
with symplectic form 
$d{\bf B }\cdot \wedge d {\bf H} + d{\bf D} \cdot \wedge  d {\bf E}$
and a Lagrangian 
submanifold which defines the constitutive relations.
If one wishes one may pass to a 13 dimensional contact manifold
with contact form $dL-{\bf D}\cdot d {\bf E} +{\bf H} \cdot d{\bf E}$.
Then the constitutive relation provides a Legendre submanifold,
which of course on projection onto the $L$ coordinate
gives back  the Lagrangian submanifold.
One may instead  perform a projection onto any pair of the 12 vector
coordinates to obtain a ``Gibbs surface'' in a seven-dimensional space.
Picking for example the pair $({\bf E}, {\bf B})$
the Gibbs surface is given by
\ben
L=1-\sqrt{1-{\bf E}^2 + {\bf B} ^2 -({\bf B}\cdot {\bf E})^2 }.
\een
This is defined only in the domain 
$D\subset \bR^6$ connected to the origin
 for which $\det(g+F) <0$, that is
 ${\bf E}^2 -{\bf B} ^2 +({\bf B}\cdot {\bf E})^2 <1$. 
Geometrically for example, to find ${\bf E}$ as a function of ${\bf D}$ and ${\bf B}$  one brings up a 6-plane parallel to the ${\bf B}$ axis whose slope is given by ${\bf D}$ until it touches the Gibbs surface. The point of contact defines ${\bf E}$. If the Gibbs surface is convex there will be only one such contact point. This is illustrated in figure \ref{gibbs}.

Convexity will guarantee that all these projections are well defined
over the relevant domain and that, the surface has no folds for example as it would if the system exhibited
some sort of  hysteresis phenomenon.
For a general non-linear electrodynamic theory
the Hessian will only be positive definite over some 
domain in  $({\bf B}, {\bf D})$ space. Outside that domain
the constitutive relation is just that: a relation rather than
a function.

The components of the Hessian are just the electric permitivities
and magnetic permeabilities. They  govern the behaviour 
of small disturbances around a background. Thus the background will be stable
as long as the Hessian is positive definite.
The equations for small fluctuations will also be hyperbolic
as long as the Hessian is positive definite \cite{B18}.

\subsection{Shock Waves and Exceptionality}

In Maxwell theory, in flat space $\bE^{3,1}$, there exist traveling
wave solutions of the form
\ben
F_{\mu \nu}= f(S) F^0_{\mu \nu}, 
\een
where $f$ is an arbitrary function of it's argument, $S={\bf n}\cdot {\bf x}- v t$, and $\bf n$ is a constant unit 3-vector.
For fixed $\bf n$ these
represent a train of parallel waves moving with unit speed
in a fixed direction.  The arbitrary function $f$ allows us to pick
the profile of the wave train arbitrarily. One may even choose it to
be discontinuous. The amplitude
of the wave is constant on a family of wave surfaces $S={\rm constant}$
which correspond to a family of spacetime parallel null hyperplanes
whose intersection with any surface  of constant time 
gives a family of parallel 2-planes in $\bE^3$. Because they
move at the speed of light, wave trains cannot be brought to rest by means of a Lorentz
transformation.

\begin{figure}
\begin{picture}(0,0)(0,0)
\end{picture}   
\centering\epsfig{file=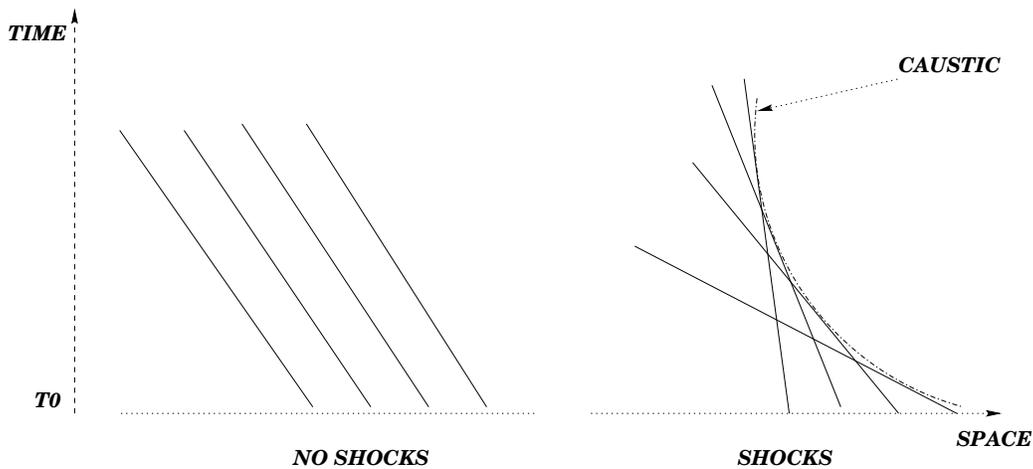,width=14cm}   
\caption{Family of hyperplanes describing the propagation of wave fronts. If the hyperplanes intersect (right figure) the theory will be singular. Regularity (left figure) arises for exceptional theories only, like Born-Infeld, which have no shock formation.}
\label{shofig}
\end{figure}

In non-linear theories
in flat space one may, by analogy, adopt   the ansatz
\ben
F_{\mu \nu}= F^0_{\mu \nu}(f(S)), 
\een   where 
$F^0_{\mu \nu}$ will now in general depend on the arbitrary function $f$ 
and where
\ben
S={\bf n}\cdot {\bf x}-v({\bf n},S)t.
\een
Now  we get a family of hyperplanes  $S=\const$ in $\bE^{3,1}$
but they are no-longer parallel, although their intersections
with any surface of constant time still
gives a family of parallel 2-planes in $\bE^3$. The wave train
therefore moves with in a constant direction but not with constant
speed. They may slow down or speed up in the sense that
a hyperplane which passes a given point in space at a later times may
have a smaller or greater speed $v({\bf n},S)$. 
The hyperplanes will thus in general intersect (see figure \ref{shofig}). At these locations
the ansatz breaks down.
Neighbouring hyperplanes will envelope  a caustic hypersurface
obtained by eliminating $S$ from the equations
\ben
S={\bf n}.{\bf x}-v(S)t, \ \ \ \ \ \ \ \ \ 1=-v^\prime(S)t,
\een      
where $^\prime$ indicates differentiation with respect to $S$. Exceptional waves are those for which
\ben
v^\prime(S)=0.
\een
If all waves are exceptional, i.e. if $v^\prime=0$ $\forall S$,
 then parallel hyperplanes are possible. If $v\ne 1$ these can be
 brought to rest by means of
a Lorentz transformation. One then has stationary solutions \footnote{We say stationary
rather than static because the Poynting vector may not vanish} depending upon
two arbitrary functions $f_1(z)$ and $f_2(z)$
of a single spatial coordinate, $z$ say. We shall give concrete examples in the next section for the Born-Infeld case.

To understand the the physical significance of exceptionality, in the sense
of the absence of shock waves, one should consider
non-exceptional theories which do admit shock waves.
As theories they are essentially incomplete. One needs
extra physical assumptions to render the evolution beyond the shock. This typically may come from some underlying
more fundamental theory. Thus the predictions
of  classical theory 
admitting shocks, or indeed other singularities,
cannot be trusted in situations where they arise or are about to arise.
In this sense such theories ``predict their own demise", 
something that is often said of classical general relativity.
By contrast a classical theory, such as Yang-Mills theory,
for which the evolution of regular finite energy initial data remains
non-singular for all times \cite{moncrief} is certainly complete
as a theory, even though, because of quantum mechanics one
does not trust every classical prediction. To check the reliability
of a classical prediction we must check to see how 
it might be effected by quantum effects.
Generally speaking, we expect classical Yang-Mills theory 
to be useful in the weak coupling limit and when dealing
with very massive excitations such as magnetic monopoles. 

Classical general relativity is known to admit singularities
as a consequence of gravitational collapse. Only for weak
data do we expect non-singular evolution  for all time \cite{christ}.
There exist fully non-linear non-singular
solutions of general relativity depending upon
two arbitrary functions propagating at unit speed.
These are the pp-waves. They may be generalized to
propagate in an Anti-de-Sitter background \cite{GRu}. In some ways AdS is analogous to a background $B$ field. But pp-waves wave-fronts in AdS are null hypersurfaces of the AdS metric. This is in contrast with 
Born-Infeld theory and other non-linear electrodynamical theories, where there are 
plane wave solutions
traveling in some non-flat background spacetime
at a slower speed. Again, by contrast with
Born-Infeld theory, the collision of two pp-waves
gives rise to a spacetime singularity \cite{Griffiths}.
Such waves definitely cannot pass through one-another.
In this respect Born-Infeld theory resembles Yang-Mills theory
more than it does general relativity \cite{moncrief}.
One is tempted to speculate that it may
be a complete classical theory. Even if this is so,
any of its  classical predictions is subject
to quantum correction unless there is some
reason, such as supersymmetry, for believing that the quantum 
corrections vanish.

\subsection{Covariant Legendre Transformation}

We introduce here a dual notation via the fields $P_{\mu \nu}$ and $N_{\mu \nu}\equiv \star P_{\mu \nu}$. This notation has the advantage of making Legendre self-duality of Born-Infeld theory manifest.

In the dual notation, the field equations of \textit{any} non-linear theory of electrodynamics are
\ben
\nabla _\mu  P^{\mu \nu}=0,
\een
or, in form language, $d\star P=0$, where  the field $P^{\mu \nu}$ is defined by
\ben
dL=-\frac{1}{2}P^{\mu \nu}dF_{\mu \nu}.
\label{dualm}
\een
$P_{\mu \nu}$ coincides with $F_{\mu \nu}$ for Maxwell's theory. In general it reads
\ben
P_{\mu \nu}= - \bigl ( L_x F_{\mu \nu} + L_y \star F_{\mu \nu} \bigr ).
\een
The components of $P_{\mu\nu}$ are just $\bf D$ and $\bf H$. Using this two-form, the energy-momentum tensor can be cast in a form identical to $T_{\mu \nu}^{Maxwell}$
\ben
T_{\mu \nu}=-P_{\mu \alpha} g^{\alpha \beta}  F_{\nu \beta} -g_{\mu \nu} L.
\een
The formulation of the theory in terms of $P_{\mu \nu}$ is dual to the $F_{\mu \nu}$ formulation in the sense of a Legendre transformation. In fact if one takes the Legendre transform with respect to $\hat L$ by
\ben
{\hat L}=-{ 1\over 2} P^{\mu \nu}F_{\mu \nu}-L,
\label{lhat}
\een
one has
\ben
d{\hat L}=-{1 \over 2} F_{\mu \nu}d P^{\mu \nu},
\een 
in analogy to (\ref{dualm}). For the special case of a purely electric configuration in flat space, $\hat{L}$ is the ordinary Hamiltonian. Introducing the Hodge dual field $N_{\mu \nu}\equiv \star P_{\mu \nu}$, and defining $s\equiv { 1\over 4} N^{\mu \nu}N_{\mu \nu}= -{ 1\over 4} P^{\mu \nu}P_{\mu \nu}$ and $t \equiv { 1\over 4} \star N^{\mu \nu}N_{\mu \nu}=-{ 1\over 4}P^{\mu \nu} \star P_{\mu \nu}$ then the theory is specified by giving  $\hat L$ as a function of $s$ and $t$. Then we have
\bequ
F_{\mu \nu}=\hat{L}_s P_{\mu \nu} +\hat{L}_t\star P_{\mu \nu}.
\label{const2}
\eequ
The energy momentum in tensor in terms of the dual variables follows from (\ref{lhat}) and (\ref{const2}):

\ben
T_{\mu \nu}=\hat{L}_s T_{\mu \nu}^{Maxwell}[P]-g_{\mu \nu}(s\hat{L}_s+t\hat{L}_t-\hat{L}), \ \ \ \  T_{\mu \nu}^{Maxwell}[P]=  -P_{\mu \alpha}g^{\alpha \beta}P_{\nu \beta} -s g_{\mu \nu}.
\een
Sufficient conditions for the dominant energy condition to hold 
are
\ben
{\hat L}_s >0, \ \ \ \ \ \ \  s{\hat L}_s+ t{\hat L}_t-\hat{L}\ge 0.
\een
In the case of Born-Infeld theory one has $t=-y$ by electric-magnetic duality invariance and expressing also $x$ in terms of $(s,t)$ one gets
\ben
-{\hat L}=1-\sqrt{1+2s-t^2} \ \ \ \ \Leftrightarrow  \ \ \ \ L(F_{\mu \nu})=-\hat{L}(N_{\mu \nu}). 
\een
For Legendre self-dual theories like Born-Infeld, the equations describing propagation of perturbations (\ref{charac}), (\ref{quadra}), will have exactly the same form in terms of the variables $(x,y)$ as they do in terms of the variables  $(s,t)$.

\sect{Born-Infeld/String theory}

\subsection{Open and Closed string metrics}

The  open string metric $G_{\mu \nu}$
is  usually obtained as follows \cite{seiwit}.\footnote{We will always use $F_{\mu \nu}$ for the gauge field, but it might represent the Kalb-Ramond potential $B_{\mu \nu}$.} 
One starts with the matrix $g+F$ whose components are $g_{\mu \nu} + F_{\mu \nu}$. Then one inverts to obtain a matrix with components
\ben
\Bigr ({ 1 \over g+F} \Bigl )^{\mu \nu}= G^{\mu \nu} + \theta ^{\mu \nu},
\een
where $G^{\mu \nu}$ is symmetric and $\theta^{\mu \nu}$ is antisymmetric.
Let $G_{\mu \nu}$ be the inverse of $G^{\mu \nu}$, i.e. $G_{\alpha \mu}G^{\mu \beta}=\delta_{\alpha}^{\beta}$. Calculation reveals that  
\bequ
G^{\mu \nu}= (G^{-1})^{\mu \nu}= \left((g-F)^{-1} g 
(g+F)^{-1}\right)^{\mu \nu}, 
\label{trouble}
\eequ
which is conformal to the inverse of (\ref{boisqu}) specialized to the Born-Infeld case. Then one checks that
\ben
G_{\mu \nu}= g_{\mu \nu}-F_{\mu \alpha}g^{\alpha \beta}F_{\beta \nu} 
\label{open}.
\een

A slightly more involved calculation shows that

\ben
\theta^{\mu \nu}=- { 1\over 1+2x-y^2} 
\bigl ( F^{\mu \nu}-y \star F^{\mu \nu} \bigr )
=- { 1\over \sqrt{1+2x-y^2} } P^{\mu \nu},
\label{noncom} 
\een
where $P^{\mu \nu}$ is the dual Maxwell field in the sense of (\ref{dualm}).
In verifying (\ref{noncom}) the following four dimensional matrix identities are useful (where {\bf 1} stands for the identity matrix):
\ben
g^{-1}F g^{-1} \star F=-y {\bf 1},
\een
\ben
g^{-1}F g^{-1}F -g^{-1}\star Fg^{-1} \star F=-2x{\bf 1},
\een
and hence
\ben
\bigl (g^{-1}F -yg^{-1} \star F\bigr )\bigl (g^{-1}F+ { 1\over y}g^{-1} \star F \bigr )=-(1+2x-y^2){\bf 1}. 
\een

Comparing (\ref{open}) with (\ref{boime}) one sees that the open string metric is equal, up to a conformal factor, to the Boillat metric governing the propagation of fluctuations around a Born-Infeld background:
\bequ
G_{\mu \nu}=\sqrt{1+2x-y^2} A_{\mu \nu}^{Boillat}.
\label{gboi}
\eequ
Relations (\ref{gboi}) an (\ref{noncom}) translate the stringy quantities $G_{\mu \nu}$ and $\theta_{\mu \nu}$ into pure non-linear electrodynamics language, i.e. the metric describing fluctuations around a fixed background and the dual Maxwell field.

An essential requirement on the causal structure defined by the open string metric is to be invariant under electric-magnetic duality rotations. To examine this we recall that the stress tensor of Born-Infeld theory, which is known to be invariant \cite{GR}, is given by
\ben
g^{\mu \nu} -T^{\mu \nu }_{\rm Born-Infeld} =
{2 \over \sqrt{-\det g}} {\partial \sqrt {-\det ( g+F)} \over \partial g_{\mu \nu}} 
\een
But 
\ben
\delta \sqrt {-\det ( g+F)}= 
{1 \over 2} \sqrt {-\det ( g+F)} \Bigl ({1 \over g+F} \Bigr ) ^{\mu \nu}
 \delta g_{\mu \nu}.
\een
Thus 
\ben
g^{\mu \nu}-T^{\mu \nu}_{\rm Born-Infeld}  ={\sqrt{-\det (g+F)} \over \sqrt{-\det g}} G^{\mu \nu} \label{tensor}.  
\een
Since the left hand side of (\ref{tensor}) is invariant so is the right hand side. Notice that the scalar $ \sqrt {-\det( g+F)} / \sqrt{-\det g} $ is not invariant but its change merely induces a conformal transformation in $G^{\mu \nu}$ and hence in $G_{\mu \nu}$, preserving the causal structure. It is worthwhile noticing that the right hand side of (\ref{tensor}) coincides with the Boillat co-metric 
\ben
g^{\mu \nu}-T^{\mu \nu}_{\rm Born-Infeld}  
 = C^{\mu \nu} _{\rm Boillat},
\label{Boillat}  
\een
which is therefore completely invariant under electric-magnetic duality rotations.
It is easily seen from (\ref{elem}) that the determinant of both sides of (\ref{Boillat}) equals $\det g^{\mu \nu}$.
Thus we get the remarkable result that
\ben
\det (\delta ^\mu_\nu - T_{\rm Born-Infeld} \thinspace ^{\mu} _{\nu} )=1.
\een

In the next subsections we illustrate the results above by analyzing the geometries seen by open strings in several special cases. We use the simplest exact solutions to Born-Infeld theory: plane wave solutions and spherically symmetric solutions.

\subsection{Exact plane wave solutions}
Boillat \cite{B14} found an exact stationary solution to Born-Infeld theory given in terms of two arbitrary functions $f_{1,2}$ of only one of the cartesian coordinates, say $z$, with an electric field and a magnetic induction given by 
\ben
\barr{c}
{\bf E}= \cosh \alpha {\bf i} + ( \cosh \alpha \sinh \beta f_1(z) -\sinh \alpha f_2(z)) {\bf k},
\\\\
{\bf B}= (\cosh \alpha f_2(z)-\sinh \alpha \sinh \beta f_1(z)) {\bf i}
-\cosh \beta f_1(z){\bf j}  + \sinh \alpha {\bf k},
\label{domwal}
\earr
\een
where $\alpha, \beta$ are arbitrary constants. The magnetic field and electric induction are easily obtained via the constitutive relations. The two Lorentz-invariants are 
\ben
-2x=1-f_1^2-f_2^2, \ \ \ \ \ \ y=f_2,
\een
so that the Born-Infeld lagrangian equals $1-|f_1|$. The Poynting
vector ${\bf P}={\bf E} \times {\bf H}$ is given by
\ben
\barr{c}
2|f_1|{\bf P}=\left(f_1^2\cosh \alpha \sinh 2\beta -2f_1f_2 \sinh \alpha \cosh \beta\right) {\bf i} +
\\\\ +\left(2f_1f_2\sinh \beta \cosh 2\alpha  -\sinh 2\alpha (f_1^2\sinh^2\beta+f_2^2+1) \right) {\bf j} -2\cosh \alpha \cosh \beta f_1 {\bf k}.
\label{poynt}
\earr
\een
One might wonder if these stationary solutions may be interpreted as domain wall solutions. That is can one choose the asymptotic values of the
arbitrary functions $f_1$ and $f_2$ so as to interpolate between
two stable ``ground states"? One would then expect to have a static family of domain walls, that is, a non trivial solution for which the Poynting vector would be zero. However, this is not allowed by $P_z$ in (\ref{poynt}): $f_1$ would need to be zero for which case ${\bf D}, {\bf H}$ blow up. Therefore one finds no domain walls, just as in Maxwell's theory.

By performing a Lorentz transformation on (\ref{domwal}), one gets the general
fully non-linear sub-luminal plane wave solution. It presents no shocks, in accordance to section 2.4., since it propagates with constant speed. In general we do not expect superposition to hold in non-linear electrodynamics and therefore such plane waves propagating on top of some background solution should not solve the equations of motion anymore. However, the plane waves obtained by boosting (\ref{domwal}) may be superimposed to a background field and still yield a solution to Born-Infeld, as shown in \cite{BB}. Therein a background magnetic field along the x-axis and electric field along the y-axis are considered: ${\bf B}= B {\bf i}$ , ${\bf E}=E{\bf j}$ so that ${\bf E} \times {\bf B}= -EB {\bf k}$. If we set
\ben
v_\pm = {-EB \pm \sqrt{1-E^2+B^2} \over 1+B^2}= { 1-E^2 \over EB \pm\sqrt{1-E^2 +B^2}}, 
\een
one checks that plane waves traveling in the z-direction can be superimposed to the background field. The waves can do so in two polarization states, with the electric and magnetic field given by:
\begin{itemize}
\item  $({\bf e}, {\bf b} ) = ( v_\pm {\bf j}, -{\bf i} )  f_\parallel (z-v_\pm t)$ for the parallel polarization state, 
\item
$({\bf e}, {\bf b} ) = ( v_\pm {\bf i}, {\bf j} )
 f_\perp (z-v_\pm t) $ for the perpendicular polarization state,

\end{itemize}
where $f_\parallel$ and $f_\perp$ are arbitrary functions of their argument. Note that there is a net drift in the $z$ direction
\ben
v_{\rm drift}=  { 1\over 2} (v_+ + v_-) = -{ EB \over 1+B^2}, 
\een
in agreement with (\ref{drift}). This drift effect may be understood
as a consequence of Lorentz-invariance.
If $B^2 >E^2$ and one performs a Lorentz boost with velocity $u=E/B$ one may
pass to a frame in which the electric field vanishes
and   the magnetic field becomes
equal to $B_0=\sqrt{B^2-E^2}$.
Now the velocity $v_0$ in the this frame is symmetric with respect to 
reversing the z-direction and is given by
\ben
v_0={ 1\over \sqrt {1+B_0^2}}.
\een
One may check that $v_\pm$, $v_0$ and $u$ satisfy
the usual relative velocity addition formula
\ben
v_\pm= {u \pm v_0 \over 1 \pm  u v_0}.
\een
If $ E^2 > B^2$ one may reduce the magnetic field to zero.
The electric field in the de-boosted frame will be $E_0=\sqrt{E^2-B^2}$
and $v_0= \sqrt{1-E_0^2}$.
In these two case the open string metrics are (using (\ref{keymetric})),
\ben
ds^2_{\rm open}=  dt^2 -dx^2 - ( 1 + B_0 ^2) ( dy^2  +dz ^2 ) 
\label{bispeb}
\een 
and 
\ben
ds^2_{\rm open}= (1-E^2_0) (dt^2-dy^2) -dx^2 -dz^2 
\een 
In terms of the electric induction $D_0$ the latter is
\ben
ds^2_{\rm open}= { 1\over 1+D^2_0}(dt^2-dy^2) -dx^2  -dz^2.
\een  
which illustrates invariance of the open  string metric up 
to a conformal factor under  the discrete electric-magnetic duality transformation
$({\bf B} , {\bf D}) \rightarrow (-{\bf D},{\bf B})$. The general metric may be obtained using a Lorentz transformation.

\subsection{BIons and other Static solutions}

Non-linear electrodynamic theories 
typically admit static finite energy solutions with distributional sources.
Because they have sources and also 
have (albeit mild) singularities, these solutions
are not solitons in the usual sense of the word. In \cite{G} they were called
BIons.
For the electrically charged BIon of Born-Infeld
theory we find the open string metric to be
\ben
ds^2_{\rm open}= { r^4 \over 1+ r^4} (dt^2-dr^2) -r^2(d\theta ^2 +\sin ^2 \theta d \phi^2 ). 
\label{BIele}
\een
The scattering of null 
geodesics is most conveniently represented as geodesics of the 
optical metric
\ben
ds_{\rm optical}^2= dr^2 + {1+ r^4 \over r^2} (d\theta ^2 +\sin ^2 \theta d \phi^2),
\een
which is easily seen to admit a 2-sphere of circular geodesics
at $r=1$ 
surrounding an infinite redshift infinite area naked singularity at 
finite proper distance situated at $r=0$. Such geodesics correspond to null geodesics of (\ref{BIele}).

The open string metric for the magnetically charged BIon is different:
 \ben
ds^2_{\rm open}= dt^2-dr^2 -{ 1+ r^4 \over r^2} (d\theta ^2 +\sin ^2 \theta d \phi^2 ). 
\een 
However the optical metrics are identical and in fact the two metrics
are conformally related.

This example may be generalized to any static configuration
in Minkowski spacetime. By static one means that the Poynting vector vanishes,
so that ${\bf E} \times {\bf B}=0$. The open string metric is
then also static and given by
\ben
ds^2_{\rm open}= (1-{\bf E}^2) dt^2 -d{\bf x}^2 +({\bf E}\cdot d{\bf x})^2
-{\bf B}^2 d{\bf x}^2 +({\bf B} \cdot d{\bf x} )^2.  
\een
Since $|{\bf E}|\le 1$ we have $G_{00}\ge 0$, moreover $G_{00}=0 $ implies
that $|{\bf E}|=1$. 
Thus any static event horizon of the Boillat metric which is not
an event horizon of the Einstein metric  must be singular, just as in the 
case of a single BIon solution.

Now consider what happens
if the closed string metric $g$ 
ceases to be flat but remains static .
One has
\ben
G_{00}= g_{00}+F_{i0} F_{j0} g^{ij}=g_{00}(1+2x).
\een
Clearly, as long as $x >-1/2$ the sign of $G_{00}$ is
determined entirely by the sign of $g_{00}$. Thus unless
the electric field reaches the critical value, there
can be no open-string static event horizon
which is not also a closed string event horizon.

This result is really obvious from the viewpoint of
electric-magnetic duality because we could instead have considered
a purely magnetic field. In this case 
\ben
G_{00}=g_{00},
\een
and the magnetic field has no effect on that part
of the metric which governs the location of event horizons. Actually, these results do not depend upon the detailed form of the
open string metric obtained from Born-Infeld theory,
nor upon electric-magnetic duality.
They hold quite generally, as may be seen directly from
the general expression (\ref{keymetric}) for the metric.
At an event horizon we need
\ben
{\bf E}^2=\zeta_\pm,
\een
That is the electric field attains it limiting value
at which point the theory breaks down.

\subsection{The Boillat metric and Bionic scattering}

In this subsection we apply the fore-going theory to the 
problem of scattering off the supersymmetric BIon or
spike solution of the Dirac-Born-Infeld equations of motion \cite{G,CM}. This has been the subject of a number of detailed studies (see \cite{kastor, park} and references therein). Physically the solution represents a fundamental string attached to a D-brane.
It is static and the transverse displacement of the brane is given by a scalar
field $\phi({\bf x})$. The metric $g$ induced on the brane
is thus 
\bequ
ds^2= dt^2 - (d{\bf x})^2 -(\nabla \phi \cdot d {\bf x} )^2. 
\eequ
Using the Bogomol'nyi conditions

\bequ
{\bf E}= \pm \nabla \phi,
\eequ
where
\bequ
\nabla ^2\phi=0, 
\eequ
the induced   metric becomes

\bequ
ds^2= dt^2 -(d{\bf x} )^2 -({\bf E} \cdot d{\bf x} )^2.
\eequ
The open string 
metric then becomes
\bequ
ds_{open}^2 =\frac{dt^2}{ 1+ {\bf E}^2} -(d{\bf x})^2.
\eequ
This metric generates the same classical scattering as the Lagrangian (77) of \cite{park}.  In the case of a single susy BIonic spike we get
\bequ
ds_{open}^2= \left({r ^4 \over 1+r^4}\right)dt^2 - dr^2 -r^2 ( d \theta ^2 + \sin ^2  \theta
d \phi ^2 ). 
\eequ
All the information about the classical scattering is now contained in $G_{\mu \nu}$.

\sect{Other spins}
Born-Infeld actions maybe extended to include scalars and spinors. In this section we shall investigate the characteristics of these fields around a background constant $B$ field. 

\subsection{Scalars}

The Born-Infeld action with a single scalar field reads
\bequ
S=\int d^4x\left(\sqrt{-det(g_{\mu \nu})}-\sqrt{-det(g_{\mu \nu}+F_{\mu \nu}-\partial_{\mu}\phi\partial_{\nu}\phi)} \right).
\eequ
Expanding around a background $F$-field and retaining only quadratic terms in $\phi$ we get the S-duality invariant expression 
\bequ
S\simeq S_{BI}+\frac{1}{2}\int d^4x \sqrt{-g}C^{\mu \nu}_{Boillat}\partial_{\mu}\phi \partial_{\nu}\phi.
\eequ
$S_{BI}$ is the usual Born-Infeld action. Therefore, as expected,  scalar perturbations propagate according to the characteristics of the open string metric. 

\subsection{Spinors}

Consider a general Dirac action of the form
\bequ
S_D=\frac{i}{2}\int d^4x  \mu  \left(\bar{\Psi}\gamma_{\alpha}a^{\alpha \beta}\nabla_{\beta}\Psi +  \dots \right),
\eequ
where $\mu$ is a scalar density, $a^{\mu \nu}=(a)^{\mu \nu}$ are the components of a contravariant second rank tensor which need be neither symmetric or antisymmetric and the ellipsis denotes other possible terms in fermions but with no derivatives. 

The gamma matrices $\gamma$ generate the Clifford algebra associated with the closed string metric $g$ 
\bequ
\{\gamma_{\alpha},\gamma_{\beta}\}=2g_{\alpha \beta}.
\eequ

The characteristics of this system are easily seen to be given by the co-metric\bequ
a^{\alpha \mu}g_{\alpha \beta}a^{\beta \nu},
\label{chara}
\eequ
which are the components of $a^t g a$. Note that we could rewrite the action as
\bequ
S_D=\frac{i}{2}\int d^4x  \mu  \left(\bar{\Psi}\Gamma^{\alpha}\nabla_{\alpha}\Psi + \dots\right) ,
\eequ
where
\bequ
\Gamma^{\alpha}=\gamma_{\beta}a^{\beta \alpha}.
\label{newgam}
\eequ
In the case of Born-Infeld theory, it is natural to take $a=(g+B)^{-1}$, in which case use of (\ref{trouble}) shows that the characteristics as determined by (\ref{chara}) are given by the open string metric. Moreover, the gamma matrices introduced in (\ref{newgam}) generate the Clifford algebra associated with the metric $a^t g a$
\bequ
\{\Gamma^{\alpha},\Gamma^{\beta}\}=2a^{\alpha \mu}g_{\alpha \beta}a^{\beta \nu}.
\eequ
For the Born-Infeld case this is the open string Clifford algebra.

Consider the Born-Infeld-Volkov-Akulov action which arises when one supersymmetrizes the Born-Infeld action. 
\bequ
S_{DBIVA}=\int d^4x\left(\sqrt{-det(g_{\mu \nu})}-\sqrt{-det(g_{\mu \nu}+F_{\mu \nu}+B_{\mu \nu}-\partial_{\mu}\phi\partial_{\nu}\phi+i\bar{\Psi}\gamma_{\mu}\nabla_{\nu}\Psi)} \right).
\eequ
In the absence of $B,\phi$ and $A$ fields this reduces to the Volkov-Akulov action \cite {VA}. Expanding to quadratic order in fermions gives a spinor action of the form $S_D$, with $\mu=\sqrt{-\det(g+B)}$ and $a=(g+B)^{-1}$. Thus, as one might have anticipated, the fermions characteristic cone is also given by the open string metric.

\subsection{Gravitons}
It is not clear whether gravitons propagating in an external $B$ field would have their characteristics modified, since these are closed string modes and propagate on the bulk. However, in the light of the fact that we now believe that gravity can be localized on the brane \cite{randsun}, one might be tempted to speculate that under some circumstances that should happen. If that is the case the obvious guess for the characteristics would be the open string metric. Indeed precisely this happens in Einstein-Schr\"odinger theory. This is a unified theory in which the usual symmetric
Einstein metric is replaced by an arbitrary $4\times4$ 
(or more generally in $n$ spacetime dimensions an $n\times n$) tensor field $a$
which we write  suggestively as:
\ben
(a^{-1}) _{\alpha \beta}= a_{\alpha \beta} =g_{\alpha \beta}+B_{\alpha \beta}.
\een
Lichnerowicz and Maurer-Tison \cite{Licher, maurer} showed that some of the small fluctuations have characteristics given 
by the symmetric part of the co-metric, i.e. ${ a } ^{(\mu \nu)}\equiv G^{\mu \nu}_{Eins-Schro}$,  in striking analogy to the open string/Born-Infeld case. Therefore the properties of these characteristics are the same as the ones presented in section 2 and 3. However, the theory exhibits a kind of bi-refringence, due to the existence of a second set of (co-)cones for small fluctuations, given by \cite{maurer}\footnote{We would like to thank M.Clayton for poining out to us the existence of this second light cone.}
\bequ
2\frac{\det{g_{\alpha \beta}}}{\det{(g_{\alpha \beta}+B_{\alpha \beta})}} g^{\mu \nu}-G^{\mu \nu}_{Eins-Schro}.
\eequ
If we define $T^{\mu  \nu}_{Eins-Schro}$ by an expression similar to (\ref{tensor}) replacing $G^{\mu \nu}$ by $G^{\mu \nu}_{Eins-Schro}$ and $F$ by $B$, our two co-metrics are conformal to, respectively:
\bequ
g^{\mu \nu}-T^{\mu  \nu}_{Eins-Schro}, \ \ \ \ \  \left(2\sqrt{\frac{\det{g_{\alpha \beta}}}{\det{(g_{\alpha \beta}+B_{\alpha \beta})}}}-1\right)g^{\mu \nu}+T^{\mu  \nu}_{Eins-Schro}.
\eequ
Just as in the BI case the first set of light cones will lie inside the Einstein cones. But because of the opposite sign in $T^{\mu  \nu}_{Eins-Schro}$, the second set of light cones will be outside both the first set of cones and the Einstein light cones. The former result was pointed out in \cite{maurer} while the latter appears to confirm the pathological properties of this theory, in that some fluctuations are tachyonic with respect to the Einstein metric. Presumably these fluctuations can carry negative energy

 We shall return to this theory in section 7. Before doing so we should recall that Einstein-Schr\"odinger theory appears to break invariance under the gauge transformation $B\rightarrow B+dA$ and for this reason it has been claimed to admit negative energy states \cite{deser}.

\subsection{Gravitinos?}
This is a short subsection because as yet we have no consistent supergravity brane solution and thus as far as we know no consistent theory of a gravitino propagating on a brane. However if such a theory exists and the gravitino propagation is affected by a background $B$ field then there is an obvious suggestion for the characteristics.

\sect{Non-Linear Electrodynamics from $U(1)$ Kaluza-Klein Theory}
Kaluza-Klein theory stems from the fact that the Ricci scalar for the $(D+1)$ dimensional ansatz 
\begin{equation}
d\hat{s}^2=ds^2+(dy+A_{\mu}dx^{\mu})^2,
\label{kklein}
\end{equation}
is $\hat{R}=-x$, i.e., the Maxwell Lagrangian. Here $ds^2$ is the D-dimensional Minkowski metric and $y$ the coordinate along the extra dimension. It is known, however, that the truncation of Kaluza-Klein theory to pure electromagnetism is not consistent. In fact, considering a trivial scalar field implies $x=constant$ via the scalar equation of motion. We will not be concerned about this point in what follows, but rather study some properties of the electrodynamical theory that arises from considering the lowest order in $\alpha'$ tree level string theory corrections to the Einstein-Hilbert action in dimensions higher than four. Full study of such Kaluza-Klein theory must be performed by considering also gravitational and scalar excitations.

With the ansatz (\ref{kklein}) the curvature invariants of second order in $D+1$ dimensions are (excluding a possible Chern-Simons term):
\begin{equation}
\begin{array}{c}
\hat{R}^2=x^2, \ \ \ \ \ \ \hat{R}_{MN}\hat{R}^{MN}=x^2+\frac{1}{2}\partial_{\beta}F_{\alpha}^{ \ \ \beta}\partial_{\mu}F^{\alpha \mu} +\frac{1}{4}F_{\mu \nu}F^{\nu \alpha}F_{\alpha \beta}F^{\beta \mu}, 
\\\\ \hat{R}_{MNPQ}\hat{R}^{MNPQ}=6x^2+\frac{5}{8}F_{\mu \nu}F^{\nu \alpha}F_{\alpha \beta}F^{\beta \mu}+\partial_{\alpha}F_{\mu \sigma}\partial^{\alpha}F^{\mu \sigma}.
\label{rsq}
\end{array}
\end {equation}
The most general parity conserving term quadratic in the curvature is then
\begin{equation}
\begin{array}{c}
\hat{R}_{MNPQ}\hat{R}^{MNPQ}+a\hat{R}_{MN}\hat{R}^{MN}+b\hat{R}^2=
\\\\ =(6+a+b)x^2+\frac{5+2a}{8}F_{\mu \nu}F^{\nu \alpha}F_{\alpha \beta}F^{\beta \mu}+\frac{4+a}{4}\partial_{\alpha}F_{\mu \nu}\partial^{\alpha}F^{\mu \nu}+...,
\label{genqua}
\end{array}
\end{equation}
where the dots stand for total derivatives. Hence, the terms with derivatives of the field strength in (\ref{rsq}) cancel (up to total derivatives) in the combination $\hat{R}_{MNPQ}\hat{R}^{MNPQ}-4\hat{R}_{MN}\hat{R}^{MN}$, thus avoiding ghosts in the propagation of the electromagnetic field. Actions with such derivative terms have nevertheless been considered in the past, as in the Bopp-Podolsky action \cite{bopp}. We will require the cancellation of ghosts and therefore consider the dimensional reduction of an action of the type
\begin{equation}
S=\frac{1}{16\pi G_{D+1}}\int d^{D+1}\hat{x} \sqrt{\hat{g}} \left(\hat{R}+\Upsilon (\hat{R}_{MNPQ}\hat{R}^{MNPQ}-4\hat{R}_{MN}\hat{R}^{MN}+b\hat{R}^2)\right).
\label{actiondp1}
\end{equation}
Specializing to $D=4$, where one can use the identity $F_{\mu \nu}F^{\nu \alpha}F_{\alpha \beta}F^{\beta \mu}=8x^2+4y^2$, we get the lagrangian
\begin{equation}
{\mathcal{L}_{KK}}= -x + \Upsilon \left((b-1)x^2-\frac{3}{2}y^2\right).
\label{lagkk}
\end{equation}
One notices the absence of an $xy$ parity breaking term to this order. In principle one could bring such term into the theory by including a Chern-Simons term. In $D+1=5$, two such possible terms are
\ben
S_{CS}=\int tr(\hat{R}_{AB}\wedge \hat{R}^{B}_{\ \ C}\wedge X), \  \ or \ \ S_{CS}=\int tr(\hat{R}_{AB}\wedge \hat{R}^{B}_{\ \ C}\wedge \hat{w}^{C}_{\ \ D}),
\een 
for some one form field $X$, or using the one form connection $\hat{w}$. The second and most natural possibility gives, however, terms of order higher than the ones considered in ${\mathcal{L}_{KK}}$. For the first possibility, the most natural choice of $X$ is as being dual to the fiber direction $\partial / \partial y$; then the first possibility contributes only to the ghosts. Hence we will not consider them anymore.

By arranging the constants $b$ and $\Upsilon$ in (\ref{lagkk}), one can recover several interesting cases which analyse in the following subsections. 

\subsection{Gauss-Bonnet electromagnetism}
This theory is obtained for $b=1$ \cite{lemos} (see also earlier work in \cite{kerner}). As pointed out in \cite{zwiebach}, an analysis for linear perturbations of the gravitational field shows that ghost cancellation requires $\hat{R}^2$ to enter the combination (\ref{genqua}) as

\begin{equation}
\Omega_{2}=\hat{R}_{MNPQ}\hat{R}^{MNPQ}-4\hat{R}_{MN}\hat{R}^{MN}+\hat{R}^2 \stackrel{\mathrm{4D}}{=}\frac{1}{4}\hat{R}_{MN}^{ \ \ \ \ AB}\hat{R}_{PQ}^{ \ \ \ \ CD}\eta^{MNPQ}\eta_{ABCD}.
\end{equation}
$\eta$ is the Levi-Civita tensor, not density. This is the Gauss-Bonnet combination. The last equality, which holds in four dimensions where the four-form $\eta$ is the volume form, shows it is the second Euler density, a topological term in four dimensions but dynamical in higher dimensions. We recall that the first Euler density, topological in two dimensions but dynamical in higher is just the Ricci scalar:
\begin{equation}
\Omega_1=\hat{R}\stackrel{\mathrm{2D}}{=}\frac{1}{2}\hat{R}_{MN}^{\ \ \ \ AB}\eta^{MN}\eta_{AB}.
\end{equation}
The last equality holds in two dimensions, where the two-form $\eta$ is the volume form. The Gauss-Bonnet combination is usually referred to as describing the first order string theory corrections to general relativity \cite{zwiebach}. The first order (in $\alpha'$) stringy gravitational action can then be written exclusively in terms of Euler densities (that does not seem to be the case already at third order):
\begin{equation}
S^{(1)}=\frac{1}{16\pi G_{D+1}}\int d^{D+1}\hat{x} \sqrt{-\hat{g}}(\Omega_1 +\Upsilon \Omega_2),
\end{equation}
with $\Upsilon \propto \alpha'$. That $S^{(1)}$ is the correct effective action relies on two arguments.  Matching the amplitude for the scattering of three on-shell gravitons in bosonic closed string theory only fixes the $(\hat{R}_{MNPQ})^2$ term; the $(\hat{R}_{MN})^2$ and $\hat{R}^2$ do not contribute to the on-shell amplitude. These are fixed by the no-ghost requirement, since one does not see any ghosts in the string spectrum. But for purely electromagnetic excitations within a Kaluza-Klein context, the no-ghost requirement is more relaxed and makes sense to consider an arbitrary $\hat{R}^2$ coefficient.
 
In non-covariant language, the Gauss-Bonnet lagrangian is described by
\bequ
\mathcal{L}_{GB}=\frac{1}{2}\left({\bf E}^2-{\bf B}^2-3\Upsilon ({\bf E} \cdot {\bf B})^2\right).
\eequ

\subsubsection{Properties of Gauss-Bonnet electromagnetism}

The constitutive relations are very simple and easily invertible. ${\bf E}$ and ${\bf H}$ may be expressed in terms of ${\bf B}$ and ${\bf D}$ as 
\bequ
{\bf E}={\bf D}+\frac{3\Upsilon({\bf B} \cdot {\bf D})}{1-3\Upsilon {\bf B}^2}{\bf B}, \ \ \ \ \  {\bf H}={\bf B}+\frac{3\Upsilon({\bf B} \cdot {\bf D})}{1-3\Upsilon {\bf B}^2}{\bf D}+\left(\frac{3\Upsilon({\bf B} \cdot {\bf D})}{1-3\Upsilon {\bf B}^2}\right)^2{\bf B}.
\eequ
It follows from the constitutive relations that ${\bf B} \cdot {\bf E}\neq  {\bf D} \cdot {\bf H}$. Therefore this theory does not admit electric-magnetic duality.

The Hamiltonian becomes
\bequ
\mathcal{H}_{GB}=\frac{1}{2}\left({\bf E}^2+{\bf B}^2-3\Upsilon ({\bf E} \cdot {\bf B})^2\right)=\frac{1}{2}\left({\bf D}^2+{\bf B}^2+\frac{3\Upsilon({\bf B}\cdot {\bf D})^2}{1-3\Upsilon{\bf B}^2}\right).
\eequ
The dominant energy condition for the theory is obeyed if $\Upsilon<0$. Hence the first expression for the Hamiltonian shows the energy is positive, whereas the one in terms of the canonically conjugate variables ${\bf B}$ and ${\bf D}$ imposes no upper-bound on the magnitude of the magnetic induction. However, the expressions for ${\bf B}$ and ${\bf D}$ in terms of ${\bf E}$ and ${\bf H}$,
\bequ
{\bf B}={\bf H}-\frac{3\Upsilon({\bf E} \cdot {\bf H})}{1+3\Upsilon {\bf E}^2}{\bf E}, \ \ \ \ \  {\bf D}={\bf E}-\frac{3\Upsilon({\bf E} \cdot {\bf H})}{1+3\Upsilon {\bf E}^2}{\bf H}+\left(\frac{3\Upsilon({\bf E} \cdot {\bf H})}{1+3\Upsilon {\bf E}^2}\right)^2{\bf E},
\eequ
\textit{do} constrain the value of the electric field to be bounded by
\bequ
{\bf E}^2=\frac{1}{3|\Upsilon|}.
\label{limite}
\eequ
Another way to see this is by using our analysis of section 2. For the Gauss-Bonnet electromagnetic theory (\ref{quadra}) becomes
\bequ
\mu =x-\frac{1}{3\Upsilon},
\label{quadragb}
\eequ
from where we can read immediately the limiting field value (\ref{limite}), in agreement with  the discussion following (\ref{quadra}). What happens to the light cones in this limit? Considering ${\bf B}=0$, we see from (\ref{velo}) that the Boillat light cone collapses in the two non-principal directions, manifesting the breakdown of the theory. Moreover, beyond such limit, the causality inequality (\ref{bradyon}) is violated.

Yet another manifestation of the limiting electric field can be seen by studying the convexity of the Hamiltonian function, as discussed in section 2.3. The latter property is equivalent to the positive-definiteness of the following six dimensional quadratic form (of the variables ${\bf b}$, ${\bf d}$):
\bequ
\barr{l}
\displaystyle{{\bf b}^2 + {\bf d}^2+ \frac{3\Upsilon}{1-3\Upsilon {\bf B}^2}\left[({\bf D} \cdot {\bf b})^2+ ({\bf B} \cdot {\bf d})^2+2({\bf b} \cdot {\bf d})({\bf D} \cdot {\bf B})  +2({\bf D} \cdot {\bf b})({\bf B} \cdot {\bf d})\right] +\frac{27\Upsilon^3}{(1-3\Upsilon {\bf B}^2)^3}\times } 
\\\\ \displaystyle{\times ({\bf b} \cdot {\bf B})^2({\bf D} \cdot {\bf B})^2 +     \frac{9\Upsilon^2}{2(1-3\Upsilon {\bf B}^2)^2}({\bf D} \cdot {\bf B})\left[4({\bf D} \cdot {\bf b})({\bf B} \cdot {\bf b})+4({\bf B} \cdot {\bf d})({\bf B} \cdot {\bf b})+({\bf D} \cdot {\bf B}){\bf b}^2\right]  .}
\earr
\eequ
Again, for vanishing magnetic induction both eigenvalues will be positive if and only if the electric field is smaller than (\ref{limite}).

Since (\ref{quadra}) reduces to (\ref{quadragb}), which has a unique solution for $\mu$ one might think that Gauss-Bonnet electromagnetism admits no bi-refringence. However, as discussed in section 2, when $\varpi$ in (\ref{quadra}) vanishes, the information contained in (\ref{quadra}) might be incomplete. As shown in \cite{lemos} this theory exhibits bi-refringence, with one cone given by the Boillat cone with (\ref{quadragb}) and the second coinciding with the Minkowski light cone. So, the middle illustration in Figure \ref{cones1} is the one to bear in mind, but now, the Minkowski light cone is degenerate; it represents both the background geometry and one of the effective geometries describing the propagation of fluctuations.

\subsection{The Born-Infeld theory to second order}
The \textit{Born-Infeld} case, ${\mathcal{L}_{KK}}={\mathcal{L}_{BI}}^{(2)}$, $b=-1/2$. The action matches the Born-Infeld action to second order. To match the constants $\Upsilon$ with $\beta$ one must remember that in the Kaluza-Klein ansatz one should replace $A_{\mu}\rightarrow \zeta A_{\mu}$ where $\zeta$ is a constant with dimension length (we are using quantum units, i.e., $c=\hbar=1$). If we write down ${\mathcal{L}_{BI}}$ as
\begin{equation}
{\mathcal{L}_{BI}}=\frac{1}{\beta^2}\left(\sqrt{-g}-\sqrt{-det(g_{\mu \nu}+\beta F_{\mu\nu})}\right),
\end{equation}
the constants match as 
\bequ
\beta^2=-\frac{3\Upsilon \zeta^4 L}{16\pi G_5},
\eequ
where $L$ is the perimeter of the compact dimension (constant since we considered a trivial dilaton).

\subsection{The Euler-Heisenberg action}
The \textit{Euler-Heisenberg} case, ${\mathcal{L}_{KK}}={\mathcal{L}_{EH}}$, $b=1/7$. This is the effective action to QED due to one-loop corrections \cite{heisen}. The constant $\Upsilon$ should then be
\begin{equation}
\frac{\Upsilon \zeta^4 L}{16\pi G_5}=-\frac{28\alpha^2}{135m_{e}^4},
\end{equation}
were $\alpha$ is the fine structure constant.

\sect{Metric independence of Black Hole properties}
In this section we consider the propagation of fluctuations of fields (including the electromagnetic) in curved backgrounds. Our main theme will be that even though there maybe more than one metric present in theory, many properties of black holes and their thermodynamic behaviour are metric invariant. In this sense we find that the event horizon and its properties have a universality which goes beyond the universality implied by the equivalence principle.  

\subsection{Causality and the Strong Energy Condition}

The presence of gravity as a background field is expected to induce changes in the propagation of electromagnetic fluctuations, in the same way a background electromagnetic field does. In fact the former is a consequence of the latter via the Einstein equations. Let us start by using the Boillat metric presented in section 2 to ask when such propagation is causal. The Boillat co-metric (\ref{boicome})
is conformal to
\ben
\left((\mu -x)L_x-yL_y+L\right)g^{\mu \nu} -\left( T^{\mu \nu} -
{ T\over 2} g^{\mu \nu} \right).
\een
From now on in this section  we assume that $g$ is the Einstein metric,
rather than some conformal multiple, such as the closed string metric.
This is because we wish to assume 
that the  Einstein equations hold.
Then  the  Boillat co-metric is conformal to
\ben
g_{\rm effective}^{\mu \nu}= g^{\mu \nu} + A R^{\mu \nu},\label{effective}
\label{geff}
\een
where $R^{\mu \nu} $ is the Ricci tensor and
\ben
{1\over 8\pi G_N A}= (x-\mu) L_x+yL_y-L.
\een
If $T_{\mu \nu}$ satisfies the strong energy condition and the Einstein equations hold, then
$R^{\alpha \beta}p_\alpha p_\beta \ge 0$ for all co-vectors
lying inside or on the Einstein co-cone. Thus if $A\ge 0$, the
 Boillat co-cone
lies outside or on the Einstein co-cone. Passing back to the original
Einstein and Boillat cones, remembering that duality reverses inclusions
we see that the strong energy condition together with the requirement
that $A \ge0 $ is a sufficient condition that the Boillat cone
lies inside the Einstein cone. In these circumstances  small
disturbances travel no faster than gravitons. 

\subsection{Stationary Event Horizons and the Touching Theorem}

Before discussing the even horizon given by the co-metric (\ref{geff}) it seems worth recalling that quantum mechanical effects renormalize the propagation equations in a background gravitational field. For scalars $\phi$ and spinors $\psi$, additional terms appear in the effective action of the form
\bequ
\frac{A}{2}R^{\alpha \beta}\partial_{\alpha}\phi \partial_{\beta}\phi, \ \ \ \ \ \  \frac{iA}{2}R^{\alpha \beta}\bar{\psi}\gamma_{\alpha}\nabla_{\beta}\psi,
\eequ
for some coefficients $A$. 
In the case of scalars these give an effective metric of the form (\ref{geff}). In the case of spinors the discussion given in section 4.2 applies. In the notation used there one has
\bequ
a^{\alpha \beta}=g^{\alpha \beta}+\frac {A}{2}R^{\alpha \beta}
\eequ
and from equation (\ref{chara}) it follows that 
\bequ
g^{\mu \nu}_{effective}=g^{\mu \nu}+AR^{\mu \nu}+\frac{A^2}{4}R^{\mu}_{ \ \alpha}R^{\alpha \nu}.
\eequ
In perturbative calculations one neglects the last third term. The second was computed by Ohkuwa within the Weinberg-Salam model \cite{ohk} yielding
\bequ
A=-\frac{11}{192\pi^2}\frac{e^2 \hbar}{M_{W}^2\sin^2\theta_W c^3},
\eequ
where $\theta_{W}$ is the Weinberg angle and $M_{W}$ is the $W$-boson mass.\footnote{Notice that the different sign in \cite{ohk} is due to the opposite convention for the Riemann tensor.} Since $A$ is negative the effective cones lie outside the Einstein cone. Physically however it is not clear that this implies the neutrino speeds faster than light, because the approximation of retaining only first derivatives in the effective action may break down.

The case of photons is more complex and it involves the Riemann tensor \cite{drum}.\footnote{It maybe of interest to note that if one has as many scalar as spinor degrees of freedom with the same mass going around the loop then the Drummond-Hathrell correction vanishes.}

Work on the causality properties of such effective metrics (\cite{shore} and references therein) uncovered a striking result  which is also relevant in the context of non-linear electrodynamics.

If the Einstein metric $g$ contains a stationary event horizon
$\cal H$ with null generators $l_\alpha$ and the weak energy condition holds,
$T^{\alpha \beta} l_\alpha l_\beta \ge 0$, then Hawking has shown that
restricted to $\cal H$ 
\ben
R^{\alpha \beta} l_\alpha l_\beta=0.
\een
It follows that 
\ben
g_{\rm effective}^{\alpha \beta}l_\alpha l_\beta=0.
\een
Thus the null generator of the horizon
lies on the effective co-cone. Passing to
the dual space we see that the Einstein cone and the effective
cone actually touch along the null generator of the horizon.
In the case that $A\ge 0$ the effective cone will touch from the inside.
This makes the existence of another effective event horizon outside
the Einstein event horizon unlikely.

\begin{figure}[t]
\begin{picture}(0,0)(0,0)
\end{picture}   
\centering\epsfig{file=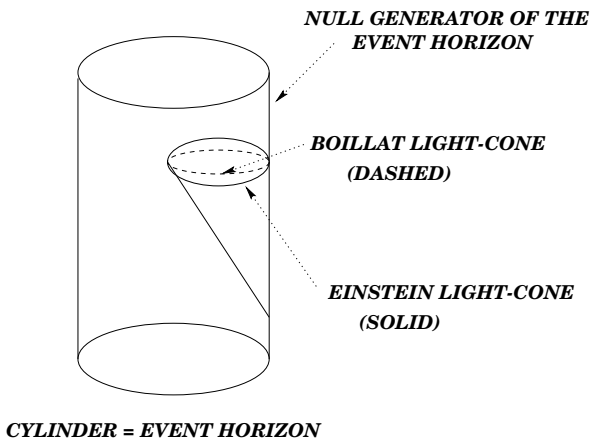,width=8cm}   
\caption{The touching theorem.}
\end{figure} 

This `Touching Theorem' shows that the concept of an absolute event horizon is more `absolute' than one might have thought. After all because quantum fluctuations will in general affect different particles differently and because the effective metric they see clearly depends upon their couplings one might have imagined that in quantum theory different particles would have different effective event horizons, in contradiction with the classical equivalence principle. However we have seen that to the order we have been working this is not so. All particles see the same event horizon. In other words, the concept of a black hole remains universal in the quantum theory.

\subsection{The surface gravity and the universality
of the Hawking temperature}

As well as the location of the event horizon one might ask whether the thermodynamic properties, such as the temperature, are universal. Because more than one metric is involved, this is not immediately obvious. 
In the case of static solutions, the simplest way of obtaining
the surface gravity $\kappa$ and hence the Hawking temperature 
$T_H= {\kappa/( 2 \pi)}$ is by setting $t=\sqrt{-1} \tau$, $\tau$ real 
and calculating the period $\beta= (T_H)^{-1} = { 2 \pi / \kappa}
$ required to remove
the potential conical singularity at the horizon.
It is clear that there will be no
conical singularity in one metric if and only if there is no conical
singularity in the other metric. Thus we get the same period
$\beta$ for both metrics.

If the timelike
Killing vector, which is of course a Killing vector
of both  metrics, is normalized to have unit magnitude at infinity
with respect the Einstein metric, 
then this calculation yields the temperature in Einstein units
as judged by closed observers at infinity. If a background
dilaton $\Phi$ is non-zero then this must be rescaled to get the 
temperature in closed string units. Similarly if the background
Kalb-Ramond field is non-vanishing we must rescale to get the temperature
in open string units. For previous work on the universality of the thermodynamic properties of black holes in generally covariant theories including arbitrary higher derivative interactions see \cite{jacobson}.

\subsection{Black Hole in a magnetic Field in Einstein-Maxwell theory}

One stimulus for this work is the current activity on physics in an external $B$ field. It is worth recalling therefore the properties of a black hole immersed  in an external magnetic field according to Einstein-Maxwell theory. The main point we wish to make is that the thermodynamic properties of the black hole are unaffected by the magnetic field passing through it. This is perhaps not unexpected if one believes that the thermodynamics has its origin in microscopic degrees of freedom whose number and nature are essentially unchanged by external fields. To be concrete the metric is \cite{ernst}
\bequ
ds^2=\Lambda(r,\theta)^2\left(-(1-\frac{2M}{r})dt^2+\frac{dr^2}{1-\frac{2M}{r}}+r^2d\theta^2\right)+\Lambda(r,\theta)^{-2}r^2\sin^2\theta d\phi^2,
\eequ
where $M$ is the analogue of the ADM mass for asymptotically Melvin boundary conditions and
\bequ
\Lambda(r,\theta)=1+\frac{1}{4}B_0^2r^2\sin^2\theta.
\eequ
The Hawking temperature $T_H$ and the area of the event horizon $A_H$ are easily seen to be the same as for the Schwarzchild solution. For some other comments on `non-commutative black holes' see \cite{garhas}.

\sect{Einstein-Schr\"{o}dinger Theory}

It is well known that there are many similarities between Born-Infeld theory and the Einstein-Schr\"odinger theory of gravity. In section 4.3 we discussed that the characteristics and therefore the causal structure relevant for fluctuations is analogous to the one for the open string. We now specialize the discussion to some black hole solutions found by Papapetrou.

The connection in this theory is not the usual Levi-Civita connection, but rather computed from the relation
\bequ
a_{\alpha \beta, \mu} - a_{\nu \beta}\Gamma^{\nu}_{\alpha \mu}-a_{\alpha \nu}\Gamma^{\nu}_{\mu \beta}=0.
\label{connec}
\eequ
The notation is the one of section 4.3. The Ricci tensor is computed by an expression formally identical to the one in General Relativity, but has both a symmetric piece $R_{(\alpha \beta)}$ and an antisymmetric one $R_{[\alpha \beta]}$. In analogy to the dual Maxwell tensor introduced in section 2.5 we define $P^{\mu \nu}= (a)^{[\mu \nu]}$. The vacuum field equations read
\bequ
R_{(\alpha \beta)}=0, \ \ \ \ \partial_{\beta}(\frak{P}^{\alpha \beta})=0, \ \ \  R_{[[\alpha \beta], \mu]}=0.
\eequ
The contravariant second rank tensor density $\frak{P}^{\mu \nu}=\sqrt{-\det{a^{-1}}} P^{\mu \nu}$. These are similar to the usual Einstein equations, equation of motion and Bianchi identities in non-linear electrodynamics, provided one thinks of $F_{\mu \nu}$ as being analogous $R_{[\mu \nu]}$. We will avoid the issue of positivity of energy in this theory.

It is perhaps worth remarking that every Ricci flat K\"{a}hler metric including Calabi-Yau spaces, provides a Euclidean solution to this theory. In fact, if $g$ is K\"{a}hler, choosing for $B$ a multiple of the K\"{a}hler form, which is covariantly constant, the Levi-Civita connection of $g$ will solve (\ref{connec}). Hence all equations of motion are obeyed. For these solutions $G^{\mu \nu}_{\rm Eins-Schro} \propto g^{\mu \nu}$ and so there is no ambiguity as to which metric to use. For the analogous phenomenon in Born-Infeld theory see \cite{garhas}.

The last comment is not true in general. Spherically symmetric solutions were found by Papapetrou \cite{papape}, corresponding to electrically' and `magnetically' charged spherically symmetric objects. The most general electrical solution reads
\ben
(a^{-1})_{\mu \nu}dx^{\mu}dx^{\nu}= ( 1+ {Q^2 \over r^4}) ( 1-{2MG_N \over r}) dt^2 -{dr^2 \over
 (1-{2MG_N \over r})} - r^2( d\theta ^2 +\sin ^2\theta d\phi^2)
 +{Q \over r^2} dt \wedge dr.
\een
A short calculation reveals that
\ben ds^2_{\rm Eins-Schro}=  
(1-{ 2G_NM \over r} ) dt^2 -{dr^2 \over (1+ {Q^2 \over r^4} )(1-{2G_NM \over r}) } - r^2( d\theta ^2 +\sin ^2\theta d\phi^2).
\een

It is striking that some of our previous findings concerning the invariance under the change of metric of the black hole properties still hold in this theory.  For example, in general the causal structure of $g$ and $G^{\rm Eins-Schro}_{\mu \nu}$  differ
but both agree about the location of the event horizon $r=2G_NM$
and its surface gravity which is 
\ben
\kappa= { 1\over 4G_NM } (1+ {Q^2 \over 16 G_N M^4} ).
\een
Note, while the area of the event horizon is given by the same formula in terms of the mass as it is in the Schwarzchild solution, a black hole with $Q\neq 0$ is hotter than the Schwarzchild hole
with the same mass. The hotter temperature is ascribable to the fact that the factor $ (1+ Q^2 / r^4) $ in $g_{00}$
 is blue-shifting rather than redshifting.

A magnetic solution found by Papapetrou reads
\bequ
(a^{-1})_{\mu \nu}dx^{\mu}dx^{\nu}=( 1-{2MG_N \over r}) dt^2 -{dr^2 \over
 (1-{2MG_N \over r})} - r^2( d\theta ^2 +\sin ^2\theta d\phi^2)+B_0 r^2 \sin\theta d\theta \wedge d\phi.
\label{papab}
\eequ
The Einstein-Schr\"odinger metric then becomes
\bequ
ds^2_{Eins-Schro}=( 1-{2MG_N \over r}) dt^2 -{dr^2 \over
 (1-{2MG_N \over r})} - (1+B_0^2)r^2( d\theta ^2 +\sin ^2\theta d\phi^2).
\eequ
The physical meaning of the two form in (\ref{papab}) is not clear. It is spherically symmetric and of constant magnitude. It is striking that the metric has a similar form to that of (\ref{bispeb}). The location of the horizon and the surface gravity are independent of $B_0$ and indeed the $r-t$ metric is identical to that of a Schwarzchild black hole.

\section{Conclusions}
The background geometry determined by a gravitational theory might not be the relevant one seen by fluctuations of some test field. This is true even at the classical level, but quantum effects can also renormalize the geometry describing the propagation of fluctuations. One quite interesting example of such distinction was uncovered in work on string propagation in a background $B$ field \cite{seiwit}: in this setup open and closed string fluctuations move, in general, at different velocities. Gravitons and Born-Infeld photons see different light cones. The discussion in sections 2 and 3 shows that the  latter causal structure is the Boillat causal structure, studied long ago in the context of non-linear electrodynamics. Moreover, the $\theta^{\mu \nu}$ parameter describing the non-commutativity of spacetime in the duality established in \cite{seiwit} is just the dual Maxwell tensor of Born-Infeld theory.

The open string metric is intrinsically connected to the Born-Infeld action, as we showed in section 4 by including scalars and fermions in a Born-Infeld type action, and showing the characteristics are determined by the open string metric. At this point a question requires more thorough understanding. In the context of string theory, the Born-Infeld action describes brane dynamics. The brane world scenario motivated by \cite{randsun} tries to bind gravitons to the brane. The difficulty is, of course, that gravitons are closed rather than open string modes. But if this program is successful, either the brane graviton sees different light cones from the other spin brane fields or, if it is governed by the open string metric, the question arises to what effective field theory describes such gravitons. The Einstein-Schr\"odinger theory seems to have exactly the characteristics we would then be looking for. But it seems to suffer from instabilities \cite{deser}.

In section 5 we looked at higher order gravity and Kaluza-Klein theory. To lowest order in $\alpha'$, the abelian truncation of the effective open string theory (Maxwell's theory) is obtained by Kaluza-Klein compactification of the effective closed string theory (Einstein's gravity). But this does not seem to hold to the next order in $\alpha'$: the Gauss-Bonnet contribution to the effective closed string theory gives upon Kaluza-Klein reduction what we named as `Gauss-Bonnet electromagnetism', distinct from both the Euler-Heisenberg theory and the Born-Infeld theory to this order.\footnote{We remark that the Born-Infeld theory to second order does not coincide with the Euler Heisenberg theory, which might be puzzling. But at least in the supersymmetric case they do coincide to this order \cite{GR}.} We remarked however that for purely electromagnetic excitations, the no-ghost requirement is weaker and by an appropriate choice of couplings one can obtain the latter two theories to this order.  It would be interesting to consider also the full Kaluza-Klein theory, with all excitations present. This would give non-minimal gravitational-electromagnetic couplings, violating the equivalence principle. Gravitational bi-refringence and dispersion effects might also be present, although the latter seem only to occur at even higher order in $\alpha'$ \cite{myers}.

In section 6 we made use of a result know in the literature as the `Touching Theorem' to show that the propagation of fluctuations in non-linear electrodynamics coupled to gravity will see a universal event horizon and black hole temperature. Such comment also holds for one-loop corrected propagators in curved spacetime. In this way black holes don't seem to `leak'. We also noticed that similar invariance is seen for a black hole immersed in a magnetic field in Einstein-Maxwell theory. It would perhaps be interesting to explicitly look at such black holes in a `Melvin Universe' for the case of non-linear electrodynamics.

\section *{Acknowledgments}
G.W.G. thanks Michael Green, Koji Hashimoto and Richard Kerner for helpful discussions. We would also like to thank M.Clayton and T.Jacobson for helpful correspondence. Some of the ideas in this paper surfaced long ago in discussions with  Malcolm Perry. C.H. is supported by FCT (Portugal) through grant no. PRAXIS XXI/BD/13384/97. This work is also supported by the PPARC grant PPA/G/S/1998/00613.

\end{document}